\documentclass[twocolumn,showpacs,superscriptaddress,prl,amsmath,amssymb,floatfix]{revtex4}
\usepackage{amsmath,amssymb}
\usepackage{bm}
\usepackage{epsfig}
\usepackage{psfrag}

\newcommand{\bd}{\bm}

\renewcommand{\onlinecite}{\cite}

\begin{document}

\title{Renormalization of  the BCS-BEC crossover by order parameter fluctuations
 }

 \author{Lorenz Bartosch}
 \affiliation{Institut f\"{u}r Theoretische Physik, Universit\"{a}t
   Frankfurt,  Max-von-Laue Strasse 1, 60438 Frankfurt, Germany}

 \author{Peter Kopietz}
 \affiliation{Institut f\"{u}r Theoretische Physik, Universit\"{a}t
   Frankfurt,  Max-von-Laue Strasse 1, 60438 Frankfurt, Germany}

 \author{Alvaro Ferraz}
 \affiliation{
 International Center for Condensed Matter Physics,
 Universidade de  Bras{\'\i}{}lia, 70910-900   Bras{\'\i}{}lia, Brazil}




 \date{July 15, 2009}

 \begin{abstract}
{
We use the functional renormalization group approach with partial bosonization in the particle-particle channel to study the effect of order parameter fluctuations on  the
BCS-BEC crossover of  superfluid fermions in three dimensions. Our approach is based on a new truncation of the vertex expansion where the renormalization group  
flow of bosonic two-point functions is closed by means of Dyson-Schwinger equations and the superfluid order parameter is related to the single particle gap via a Ward identity. 
We explicitly calculate the chemical potential, the single-particle gap, and the superfluid order parameter
at the unitary point and compare our results with experiments and previous calculations.
}
\end{abstract}


\maketitle

The BCS-BEC crossover in a two-component Fermi gas
has attracted the attention of theorists for several
decades \cite{Eagles69,Leggett80,Nozieres85,Drechsler92,Randeria95}.
It is generally accepted that the nature of  the superfluid state exhibits
a smooth crossover as a function of the dimensionless 
parameter $  1/  k_F a_s $, where $k_F$ is the Fermi 
momentum and $a_s$ is the $s$-wave scattering length in vacuum.
While for a small negative scattering length, i.e., $ 1/ k_F a_s \ll -1$,
the paired state generated by the attractive interaction
is a collection of spatially extended Cooper pairs 
(BCS limit), in the opposite limit
$ 1/ k_F a_s \gg 1$ 
 the superfluid state can be viewed as a Bose-Einstein condensate
of tightly bound fermion pairs (BEC limit).
Of particular interest is the unitary point  $1/ k_F a_s =0$ where
the scattering length diverges and
 $k_F$ sets the only length scale of the system.
In this regime
quantitative calculations are difficult because there is no small parameter 
to justify approximations.

In the past few years several observables
such as the chemical potential and the quasi-particle gap 
have been determined experimentally at the unitary 
point~\cite{Bartenstein04,Bourdel04,Kinast05,Partridge06,Bloch08},
but there is still some uncertainty in the
precise numerical values of these quantities.
The unitary point has also been studied
theoretically using Monte Carlo simulations~\cite{Carlson03,Astrakharchik04} 
and various analytical methods based on
field theoretical techniques~\cite{Nishida06,Veillette07,Nikolic07,Haussmann07,Diener08} 
or
the functional renormalization group 
(FRG)~\cite{Krippa07,Diehl07,Floerchinger08,Strack08},
but also in this case the theoretical results have not converged yet.
In such a situation it is desirable to study 
this problem using new approximation strategies which are complementary to
previous calculations. 
In this work we shall therefore develop a novel  FRG approach
for the BCS-BEC crossover which is based on
a suitable truncation of the vertex expansion 
using skeleton equations and Ward identities.
For fixed density we explicitly calculate the chemical potential,
the single-particle gap, and the superfluid order-parameter
%
%
at the unitary point in three dimensions, and compare our 
results with experiments and with previous calculations.

We consider a system of neutral fermions with energy dispersion 
$\epsilon_{\bd{k}} = \bd{k}^2  / (2m)$ and a short-range  attractive
two-body interaction $g_{\bd{p}}$ depending  on the total momentum $\bd{p}$ 
of a fermion pair. 
After decoupling
the interaction in the particle-particle channel using a
complex Hubbard-Stratonovich field $\chi$,  the Euclidean
action of our model can be written as $S = S_0 + S_1$, with Gaussian 
part $S_0$ and interaction $S_1$ given by \cite{footnotelabels}
 \begin{eqnarray}
 S_0 & =  & \sum_ {\sigma} \int_K (  - i \omega +  \xi_{ \bd{k}  })
 \bar{\psi}_{K \sigma} \psi_{K \sigma} +  \int_P  g^{-1}_{\bd{p}} \bar{\chi}_{P} \chi_P ,
 \label{eq:S0def}
 \\ 
 S_1
 & = &   \int_P \int_K [ 
 \bar{\psi}_{ K + P  \uparrow} \bar{\psi}_{ -K   \downarrow}  {\chi}_P 
 +   {\psi}_{ -K   \downarrow}  {\psi}_{ K + P  \uparrow} \bar{\chi}_P   ].
 \label{eq:S1}
 \end{eqnarray}
Here, the energy $\xi_{\bd{k}} =
\epsilon_{\bd{k}} - \mu$ is measured relative to 
the chemical potential $\mu$, and the  anti-commuting fields
$\psi_{ K \sigma}$ and $\bar{\psi}_{ K \sigma}$ represent
fermions with  energy-momentum  $K=(i\omega,\bm{k})$ and 
spin projection $\sigma$.  The complex bosonic field $\chi_P$
is conjugate to the fluctuation of the superfluid order parameter with
energy-momentum $P = (i\bar\omega, \bm{p})$. For convenience
we choose our sign convention such that
$g_{\bd{p}} > 0 $ for attractive interactions.

\begin{figure}[tb]
  \centering
  \epsfig{file=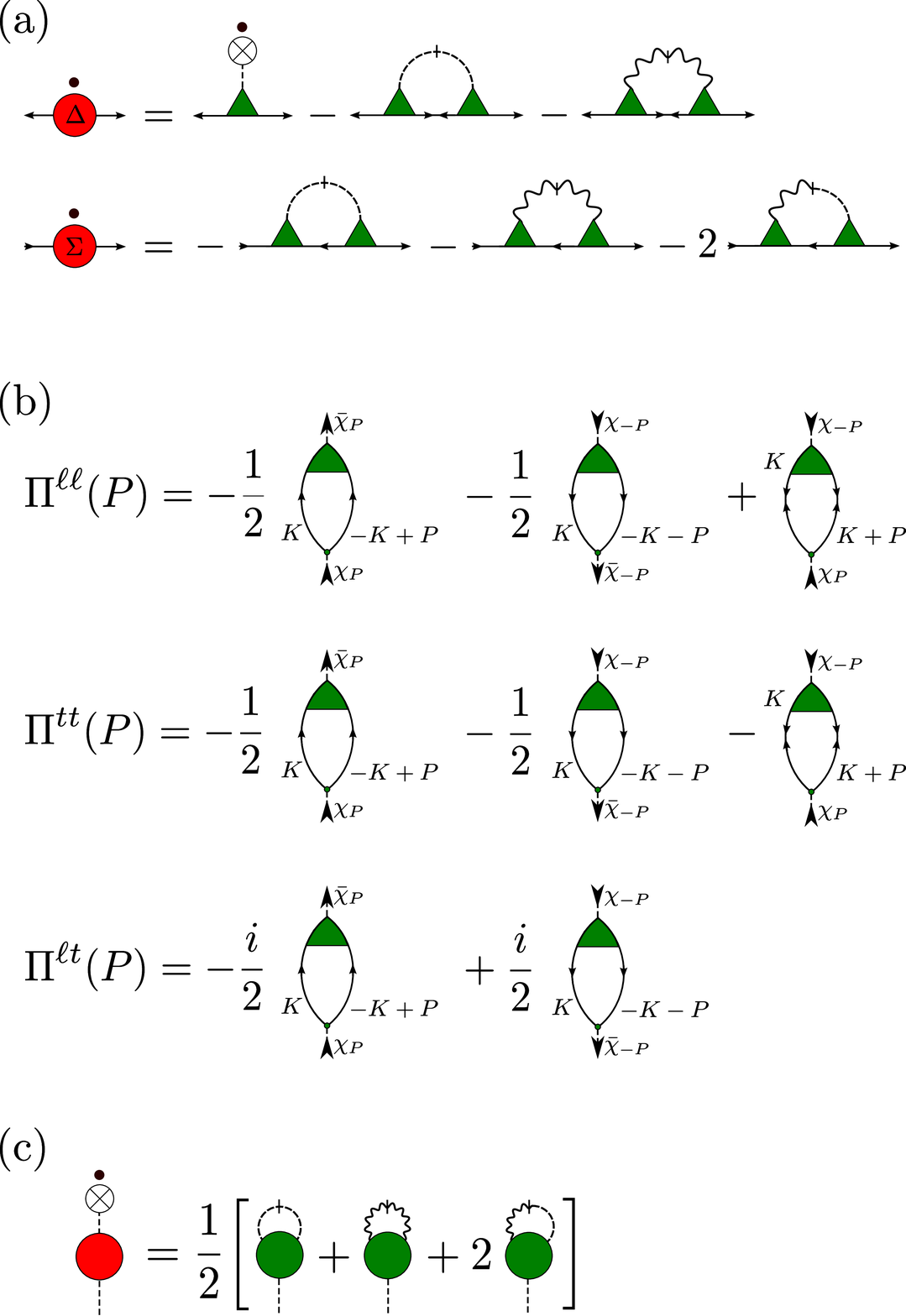,width=86mm}
  \caption{%
(Color online)
(a) Diagrammatic representation of our approximate FRG flow equations
(\ref{eq:flowselfanomal}, \ref{eq:flowselfnormal}) for the fermionic self-energies
$\Delta ( K )$ and $\Sigma ( K )$.
(b) Exact skeleton equations for the bosonic self-energies
$\Pi^{\ell \ell} ( P ), \Pi^{\ell t } ( P ), \Pi^{tt} ( P )$.
(c) Exact FRG flow equation for the order parameter $\langle \chi \rangle$.
Solid arrows represent fermionic propagators, dashed 
arrows represent the complex boson fields $\chi$ and $\bar{\chi}$,
dashed lines without arrow represent 
longitudinal components $\varphi^{\ell}$  of 
$\chi = ( \varphi^{\ell} + i \varphi^t )/\sqrt{2}$, while wavy lines
represent the transverse components $\varphi^{t}$.
}
  \label{fig:diagrams}
\end{figure}

To derive FRG flow equations for our model
 we introduce a  cutoff $\Lambda$ into the
Gaussian propagators appearing in Eq.~(\ref{eq:S0def}) and
consider the evolution of the generating functional of the irreducible vertices
as the cutoff is reduced \cite{Wetterich93,Morris94}. 
The physical vertices are then recovered
for $\Lambda \rightarrow 0$.
The FRG equations for the irreducible vertices of the above mixed
boson-fermion theory follow as a special case of
the general FRG flow equations derived in
Ref.~\onlinecite{Schuetz05}.
In contrast to a previous FRG calculation \cite{Strack08},
we use here a scheme where the cutoff is introduced only in the bosonic 
part of the Gaussian propagator.  The advantage of our 
boson cutoff scheme is that the initial condition for the
fermionic self-energy is simply given by
the self-consistent  Hartree-Fock approximation (i.e. the BCS approximation),
while the initial vertices in the bosonic sector 
are given by closed fermion loops with an arbitrary number of bosonic external 
legs \cite{Schuetz05}. 
In particular, at the initial scale the bosonic two-point functions 
are given by  the ladder approximation.
In order to obtain numerically tractable FRG equations,
we have to make further approximations: first of all,
we neglect the momentum-frequency dependence of
the vertices with one boson and two fermion legs,
replacing these vertices by  a momentum- and frequency-independent coupling 
$\gamma = \gamma_\Lambda$. Moreover, we
completely ignore vertices with two fermion legs and
more than one boson leg.
Within these approximations, the FRG flow equations for
the anomalous ($\Delta$) and normal ($\Sigma$) fermionic self-energies 
in our cutoff scheme are
 \begin{align}
  \partial_{\Lambda} \Delta ( K ) & = \gamma \partial_{\Lambda} \langle \chi \rangle 
  +
 \frac{ \gamma^2}{2}  \int_P  \bigl[ \dot{F}^{ \ell \ell }_P - \dot{F}^{tt}_P \bigr]
  {A} ( P-K )  \; ,
 \label{eq:flowselfanomal}
 \\ 
\partial_{\Lambda} \Sigma ( K ) & = 
- \frac{ \gamma^2}{2} \int_P  \bigl[ \dot{F}^{\ell \ell  }_P + \dot{F}^{ t t }_P 
- 2 i \dot{F}^{\ell t}_P 
\bigr] {B} ( P- K ) \; , 
  \label{eq:flowselfnormal}
 \end{align}
which are shown graphically in Fig.~\ref{fig:diagrams}~(a).
Here, $A ( K)$ and $B ( K )$ are the anomalous and normal 
component of the fermionic single-particle Green function,
which are related to the self-energies by 
\begin{align}
  & A ( K ) = - \frac{\Delta ( K )}{ D ( K ) } \;, \\
  & B ( K ) =  \frac{G^{-1} ( -K )}{ D ( K )}  \;, 
\end{align}
where 
$G^{-1} ( K )  = i \omega - \xi_{\bd{k}} - \Sigma ( K )$ and
$D ( K ) = G^{-1} ( K ) G^{-1} ( -K ) + | \Delta ( K ) |^2$.
The functions $\dot{F}^{ i j }_P$ in Eqs.~(\ref{eq:flowselfanomal}) and  (\ref{eq:flowselfnormal})
are the bosonic single-scale propagators associated with
longitudinal (upper index $\ell$) or transverse (upper index $t$) fluctuations of the
field $\chi$. 
Choosing for simplicity a sharp momentum cutoff in the bosonc sector,
the single-scale propagators are  
$\dot{F}^{ i j }_P = - \delta ( \Lambda - | \bd{p} | ) {F}^{ i j }_P$,
where the bosonic propagators ${F}^{ i j }_P$
can be expressed in terms of the irreducible bosonic self-energies 
$\Pi^{ ij } ( P )$ (polarizations) as
 \begin{align}
 & \left[ \begin{array}{cc}
 F^{\ell \ell}_P  & F^{ \ell t}_P  \\
F^{t  \ell }_P   & F^{tt}_P 
 \end{array} \right]
 = \nonumber \\
& \qquad \quad \frac{ 1}{   N(P)}
 \left[ \begin{array}{cc}
  g_{\bd{p}}^{-1} +   \Pi^{tt} (P ) & 
 -   \Pi^{\ell t } (P ) \\
 -  \Pi^{t \ell} (P ) & 
  g_{\bd{p}}^{-1} +   \Pi^{\ell \ell} (P ) 
 \end{array} \right] ,
 \label{eq:Fdef}
 \end{align}
with 
\begin{equation}
 N ( P )  =  [   g_{\bd{p}}^{-1} +  \Pi^{\ell \ell} (P ) ]   
 [  g_{\bd{p}}^{-1} +  \Pi^{tt} (P )  ] +   [\Pi^{\ell t} (P )]^2 
 \; .
\end{equation} 
In order to determine the fermionic self-energies
from Eqs.~(\ref{eq:flowselfanomal}) and (\ref{eq:flowselfnormal})
we need additional equations for the polarizations
$\Pi^{ ij } ( P )$ and for the flowing order parameter
$\langle \chi \rangle $. Instead of explicitly writing down the FRG flow equations
for $\Pi^{ ij } ( P )$, we shall use  skeleton equations 
(which follow from Dyson-Schwinger equations \cite{Schuetz05})
to relate the bosonic self-energies to the fermionic ones~\cite{Bartosch09a}.
Graphically, the exact skeleton equations for the irreducible bosonic self-energies
$\Pi^{ ij } ( P )$
are shown in Fig.~\ref{fig:diagrams}(b).
Within our truncation where the three-legged boson-fermion vertex is approximated 
by a flowing coupling $\gamma = \gamma_{\Lambda}$,
these relations become
  \begin{subequations} 
\begin{eqnarray}
  \Pi^{ \ell \ell} ( P ) &  = &  
- \frac{\gamma}{2} \int_K
\Bigl[ {B} ( K ) B ( - K+P ) 
- {A} ( K ) A ( K +P)  
\nonumber
 \\
  & & \hspace{13mm}
  +( P \rightarrow -P) \Bigr] \; ,
 \label{eq:Pillskel} 
 \\
  \Pi^{ tt} ( P )
  & = &
- \frac{\gamma}{2} \int_K
\Bigl[ {B} ( K ) B ( - K+P )  
+ {A} ( K ) A ( K +P)  
\nonumber
 \\
  & & \hspace{13mm}
 +( P \rightarrow -P) \Bigr] \; ,
 \label{eq:Pittskel} 
 \\
  \Pi^{ \ell t} ( P )
  & = &
 - \frac{i \gamma}{2} \int_K
 \Bigl[  B ( K ) B ( - K+P ) 
\nonumber
 \\
  & &\hspace{13mm}
  - ( P \rightarrow -P) \Bigr] \; .
 \label{eq:Piltskel} 
 \end{eqnarray}
 \end{subequations}
To close our system of flow equations, we still need an equation
for the vertex renormalization factor $\gamma$ and the
flowing order parameter $\langle \chi \rangle$
appearing in Eq.~(\ref{eq:flowselfanomal}).
In our boson cutoff scheme, the flow of    $ \langle \chi \rangle$
is driven by the irreducible vertices with three bosonic legs, 
$\Gamma^{\ell \ell \ell}, \Gamma^{\ell \ell t},\Gamma^{\ell t t}$,
as shown graphically in Fig.~\ref{fig:diagrams}(c).
The crucial point is now that 
from the requirement that the
FRG flow  preserves the gapless nature of the 
Bogoliubov-Anderson (BA) mode \cite{Schrieffer83}
we can easily obtain an expression for
$\partial_{\Lambda} \langle \chi \rangle$
without explicitly considering the RG flow of
the bosonic three-legged vertices.
Therefore we note that the condition for the vanishing of  the gap
of the BA mode is 
\begin{equation}
  g_0^{-1} + \Pi^{tt} (0) =0 \;. 
\label{eq:gaplesscondition}
\end{equation}
This condition is obviously
satisfied at the initial RG scale $\Lambda = \Lambda_0$ because
the ladder approximation with self-consistent Hartree-Fock propagators
is conserving. To make sure that the BA mode remains gapless for any value of 
$\Lambda$ we simply require that
Eq.~(\ref{eq:gaplesscondition}) remains valid during the entire RG flow.
With $\Pi^{tt} (0)$ given by Eq.~(\ref{eq:Pittskel}), this is an implicit relation between
$\partial_{\Lambda} \Delta (K )$ and $\partial_{\Lambda} \Sigma (K )$.
By demanding that this relation is consistent with 
Eqs.~(\ref{eq:flowselfanomal}) and (\ref{eq:flowselfnormal}) we can uniquely fix the
RG flow of the order parameter $\langle \chi \rangle$.
Finally, using the $U(1)$-gauge symmetry of the action given in Eqs.~(\ref{eq:S0def}) 
and (\ref{eq:S1}) we can derive a Ward identity which relates the 
ratio of the anomalous self-energy and the superfluid order parameter
to the vertex renormalization factor $\gamma$,
\begin{equation}
  \label{eq:WardIdentity}
  \Delta(0) = \gamma \langle \chi \rangle \;.
\end{equation}
Eqs.~(\ref{eq:flowselfanomal})--(\ref{eq:WardIdentity}) 
form a closed system of integro-differential
equations for the fermionic self-energies $\Delta ( K )$, $\Sigma ( K )$,
the vertex renormalization factor $\gamma$ and 
the order parameter $\langle \chi \rangle$, which should be solved
with the initial conditions $\Delta ( K )_{\Lambda_0} = \langle \chi \rangle_{\Lambda_0}
= \Delta_0$ and $\Sigma ( K )_{\Lambda_0} =0$. Here, $\Delta_0$ is the
single-particle gap in the BCS approximation.

The numerical analysis of the system of coupled integro-differential 
equations (\ref{eq:flowselfanomal})--(\ref{eq:WardIdentity}) is beyond the scope of this work.
Here, we further simplify these equations by neglecting the momentum-dependence of the
fermionic self-energies and keeping only the linear frequency correction to the normal self-energy, 
replacing $\Delta ( K ) \rightarrow \Delta$ and
$\Sigma ( K ) \rightarrow \Sigma  -(Z^{-1} -1) i \omega$, where
\begin{equation}
  \label{eq:ZBCSBEC}
  Z= \frac{1}{1- \left. \frac{ \partial \Sigma(i\omega,0) }{\partial (i\omega )} 
\right|_{\omega=0}}
\end{equation}
is the inverse flowing wave function renormalization factor.
 The flowing single-particle 
propagators are then given by  $A ( K ) = Z \tilde{A} ( K ) $ and
$B ( K ) = Z \tilde{B} ( K ) $, where
\begin{align}
 & \tilde A ( K ) = - {\tilde\Delta}/({ \omega^2 + \tilde E_{\bd{k}}^2}) \;, \\
 & \tilde B ( K )  =  - ( i \omega + \tilde\xi_{\bd{k}  } )/({ \omega^2 + \tilde{E}_{\bd{k}}^2} ) \;,
 \label{eq:Bflowdef}
\end{align}
with
  $\tilde E_{ \bd{k}  }   =  [\tilde \xi_{\bd{k} }^2 + \tilde \Delta^2 ]^{1/2}$, 
$\tilde{\xi}_{\bd{k}  }  = \tilde{\epsilon}_{\bd{k}} - \tilde \mu$,
$\tilde{\epsilon}_{\bd{k}} = Z \epsilon_{\bd{k}}$,
$\tilde \mu = Z(\mu-\Sigma) $, and
$\tilde \Delta = Z\Delta$.
It should be noted that $\tilde \Delta$ can be identified with the physical single particle gap which can be measured, e.g., in tunnelling experiments.
With these approximations, our system of flow equations reduces in $D$ dimensions to
\begin{eqnarray}
 \Lambda \partial_\Lambda \tilde \mu   & = & \eta \tilde \mu
  - \gamma
  \left( \Lambda / k_{F,0} \right)^{D} \epsilon_{F,0} \nonumber \\
& & \times  \int \frac{d\bar \omega}{2\pi} \bigl[ \tilde F^{\ell \ell }_P + \tilde F^{tt }_P - 2i \tilde F^{\ell t }_P\bigr] \tilde B(P) \;,
  \label{eq:flowtildemuBCSBEC} \\  
   \Lambda \partial_\Lambda \ln \gamma  & = & 
  - ( \gamma / \tilde \Delta ) 
  \left( \Lambda / k_{F,0} \right)^{D} \epsilon_{F,0}   
\nonumber \\
&  & \times  
\int \frac{d\bar \omega}{2\pi} \bigl[ \tilde F^{\ell \ell }_P - \tilde F^{tt }_P \bigr] \tilde A(P) \;.
  \label{eq:flowgammaBCSBEC}
\end{eqnarray}
The wave function renormalization factor is
determined by $\Lambda \partial_{\Lambda}  Z = \eta  Z$,
with the flowing anomalous dimension
\begin{eqnarray}
  \eta  & = &  \gamma
   \left( \Lambda / k_{F,0} \right)^{D} 
\epsilon_{F,0}  
 \int \frac{d\bar \omega}{2\pi} \bigl[ \tilde F^{\ell \ell }_P + \tilde F^{tt }_P - 2i \tilde F^{\ell t }_P\bigr]
\nonumber \\
& & 
\hspace{10mm} \times
\frac{\tilde E_{\Lambda}^2 -\bar \omega^2 + 2 i \bar \omega  \tilde \xi_{\Lambda}}{(\bar \omega^2 + \tilde E_{\Lambda}^2 )^2}  \;.  
\label{eq:flowetaBCSBEC}  
 \end{eqnarray}
Due to the sharp momentum cutoff we may set  
$P=(i\omega,\Lambda)$.
The FRG flow is further constrained by the condition
(\ref{eq:gaplesscondition}) that the BA mode is gapless 
and by the relation 
\begin{equation}
\tilde \Delta  = Z \gamma \langle \chi \rangle \;,
\label{eq:rescaledWardIdentity}
\end{equation}
imposed by the Ward identity (\ref{eq:WardIdentity}).
The dimensionless interaction terms $\tilde F^{ij}_P$ appearing above are 
defined by $\tilde F^{ij}_P = Z^2 \gamma \nu_0 F^{ij}_P$, and 
$\epsilon_{F,0}$, $k_{F,0}$, and $\nu_0$  denote the Fermi energy, 
the Fermi wave vector and the density of states at the Fermi energy (per spin projection) of 
a non-interacting system which has exactly the same density as our interacting
system at the initial scale $\Lambda = \Lambda_0$.

\begin{figure}[tb]
  \centering
  \epsfig{file=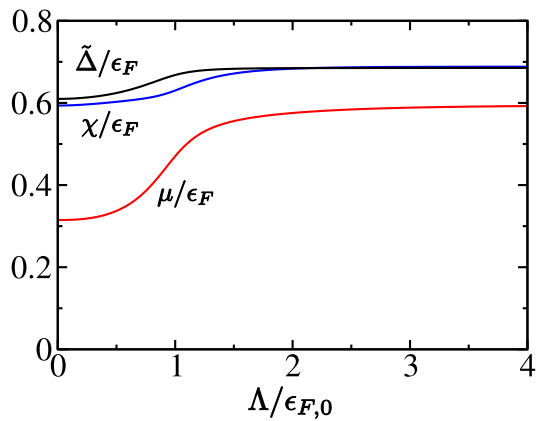,width=70mm}
\caption{%
(Color online)
RG flow the order parameter
$ \langle \chi \rangle$, the single-particle gap  
$\Delta_{\Lambda}$, and the chemical potential $\mu$ 
at the unitary point  in $D=3$. Here,
$\epsilon_F$ is the Fermi energy of a noninteracting system
which has the same density as our interacting system. 
}
  \label{fig:RGflowchiDeltamu}
\end{figure}

For convenience, we work with a
momentum-independent bare coupling
 $g_{\bd{p}} \rightarrow g_0$. In dimensions 
$D \geq 2$ this gives rise to an ultraviolet divergence
in the BCS gap equation 
which also appears in the polarizations 
$\tilde \Pi^{ii}(P) = \Pi^{ii}(P)/(Z^2 \gamma \nu_0)$.
For $D > 2$ we may absorb this divergence into the bare coupling
by introducing the dressed coupling $g $ via
$(Z^2 \gamma g)^{-1} = (Z^2 \gamma g_0)^{-1} - V^{-1} \sum_{ \bd{k}} (2 \tilde \epsilon_{\bd{k}})^{-1}$, 
so that the constraint (\ref{eq:gaplesscondition}) that the BA mode remains gapless 
turns into
 \begin{equation}
   \frac{1}{ Z^2 \gamma g} = 
  \frac{1}{V} \sum_{\bm{k}}
   \left[ \frac{1}{2\tilde E_{\bm{k}}}
     -  \frac{ 1}{ 2 \tilde \epsilon_{\bm{k}} }
   \right] .
   \label{eq:gBCSFRG}
 \end{equation}
In $D=3$ the dressed coupling is related to the
$s$-wave scattering length $a_s$ via
$g = - 4 \pi a_s /m$.
It is intriguing to note that we can derive exactly the same equation by means of a skeleton equation for the order parameter. This means that the gaplessness of the BA mode is in fact a natural consequence of our truncation scheme.
\begin{figure}[tb]
  \centering
  \epsfig{file=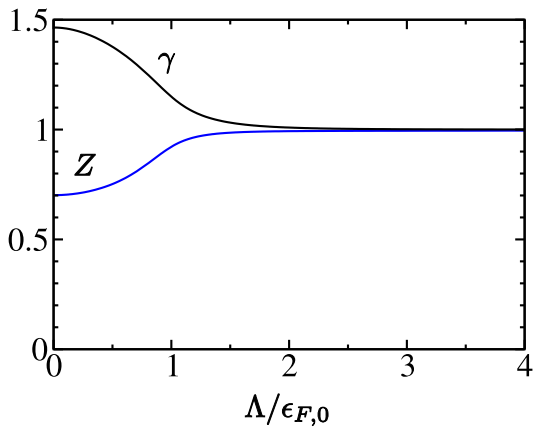,width=70mm}
\caption{%
(Color online)
RG flow the three-legged vertex $\gamma$  and the wave function renormalization factor $Z$
at the unitary point ($k_F a_s = \infty$) in three dimensions.
}
  \label{fig:RGflowZgamma}
\end{figure}
Together with the rescaled Ward identity~(\ref{eq:rescaledWardIdentity}) 
and the gapless condition~(\ref{eq:gBCSFRG}), 
the flow equations (\ref{eq:flowtildemuBCSBEC})--(\ref{eq:flowetaBCSBEC})
form a closed system of RG equations  
for  the five parameters
$ \langle \chi \rangle$, $\tilde \Delta$, $\tilde \mu = Z(\mu-\Sigma)$, $\eta$ and $\gamma$ which can be solved numerically without further 
approximation. All bosonic polarizations can be expressed in terms of these parameters via skeleton approximations.
For every value of the cutoff $\Lambda$ this requires the numerical evaluation of a three-dimensional integral.
The RG flow of the three-legged vertex $\gamma$ and the 
wave function renormalization factor $Z$  
at the unitary point in three dimensions is shown in Fig.~\ref{fig:RGflowZgamma}.
Moreover, in Fig.~\ref{fig:RGflowchiDeltamu} we present our results for
the order parameter
$ \langle \chi \rangle$, the single-particle gap  
$\tilde \Delta$, and  the chemical potential $\mu$ in
units of the Fermi energy $\epsilon_F$ of a noninteracting system 
having exactly the same density as our flowing system. 
Note that
$\epsilon_F$ is determined by the true density of the system which 
we calculate  via the normal component of the one-particle Green function
for a given value of the cutoff $\Lambda$.
At the end of the flow we find for the renormalized quantities
\begin{equation}
 \mu/\epsilon_{F} = 0.32,  \quad 
\tilde{\Delta} / \epsilon_F   =  0.61,  \quad
 \langle \chi \rangle / \epsilon_F  =  0.59 .
\label{eq:FRGunitary} 
\end{equation} 
At the unitary point we may calculate
the ground state energy per particle from
$\varepsilon_0 = 3 \mu /5 =0.19$.
The above numbers should
be compared with the mean-field results
$\mu / \epsilon_{F} = 0.59$ and $\tilde{\Delta} / \epsilon_F =  \langle \chi \rangle / \epsilon_F = 0.69$.

Our value $\mu / \epsilon_F =0.32$ is smaller than the Monte Carlo 
results $0.44$
(Ref.~\onlinecite{Carlson03}) and $0.42$ (Ref.~\onlinecite{Astrakharchik04}), 
but it is quite close to the
value $0.36$ obtained by Haussmann {\it{et al.}}\cite{Haussmann07}
and agrees perfectly with the experiment 
by Bartenstein {\it·{et al.}} \cite{Bartenstein04}. 
Our value  for the renormalized single-particle gap is within the error bars
of the Monte Carlo simulations \cite{Carlson03,Astrakharchik04}, and agrees with 
the FRG calculation by Diehl {\it{et al.}}\cite{Diehl07}, while
the FRG result for $\tilde \Delta / \epsilon_F$
 by Krippa~\cite{Krippa07} is only $3 \%$ smaller than the mean-field result.
In contrast to our approach, the FRG calculations of Refs.~\cite{Diehl07,Krippa07}
are based on a truncated gradient expansion and do
not  distinguish between the quasi-particle gap and the order parameter,
which are conceptually different quantities.


In summary, using a new truncation strategy 
of the FRG flow equations for the BCS-BEC crossover
we have calculated the chemical potential, the single-particle gap, and the
order parameter
at the unitary point in three dimensions and obtained reasonable agreement with
experiments and other calculations.
In contrast to the truncated 
derivative expansion of FRG flow equation for the BCS-BEC crossover
used in
Refs.~\onlinecite{Diehl07,Floerchinger08}, our strategy is based 
on a truncation of the vertex expansion of the partially bosonized  theory using
Dyson-Schwinger equations and Ward identities.
Moreover, by introducing a momentum cutoff only in the bosonic sector, 
we can directly calculate fluctuation corrections to the mean-field approximation,
which serves as the initial condition for the FRG flow. 
The strong vertex correction of almost $50 \%$ shown in Fig.~\ref{fig:RGflowZgamma}
shows that Eliashberg type approximations
are not quantitatively accurate close to the unitary point.

\acknowledgments
This work was  supported by the DFG via SFB/TRR49 and by a DAAD/CAPES
PROBRAL grant.


\begin{thebibliography}{99}
\bibitem{footnotelabels}
We denote by  $K = ( {\bd{k}} , i \omega )$ and $P = ( {\bd{p}} , i \bar{\omega}  )$
collective labels
for momenta and  Matsubara frequencies.
Integration symbols are
  $\int_K = (\beta V )^{-1} \sum_{ {\bd{k}}, \omega } $ and
  $\int_P = (\beta V )^{-1} \sum_{ {\bd{p}}, \bar{\omega} } $
where  $V$ is the volume and $\beta $ is the inverse temperature. The corresponding
normalization of the delta-symbols is 
$\delta_{K,K'}=\beta V
\delta_{\omega,\omega'}\delta_{\bd{k},\bd{k}'}$.
We focus on the zero-temperature limit and infinite volume limit 
throughout this work.
\end{thebibliography}
\end{document}